\documentclass{ws-ijmpa}
\usepackage[super,compress]{cite}
\usepackage{graphicx}
\usepackage{amsmath, amssymb}
\usepackage[implicit=false]{hyperref}

\begin{document}
\markboth{Ruo-Yu Liu}{The Physics of Pulsar Halos}

\catchline{}{}{}{}{}
%
%%%%%%%%%%%%%%%%%%%%%%%%%%%%%%%%%%%%%%%%%%%%%%%%%%%%%%%%%%%%%%%%%%%%

\title{The Physics of Pulsar Halos: Research Progress and Prospect}

\author{Ruo-Yu Liu}

\address{$^1$School of Astronomy and Space Science, Nanjing University\\
Xianlin Avenue 163, Nanjing, 210023, China\\
$^2$Key laboratory of Modern Astronomy and Astrophysics (Nanjing University)\\
Ministry of Education, Nanjing 210023, China\\
ryliu@nju.edu.cn}

%\author{Second Author}
%
%
%\address{Group, Laboratory, Address\\
%City, State ZIP/Zone, Country\\
%second\_author@domain\_name}

\maketitle

\begin{history}
\received{Day Month Year}
\revised{Day Month Year}
\end{history}

\begin{abstract}
Diffusive TeV gamma-ray emissions have been recently discovered extending beyond the pulsar wind nebulae of a few middle-aged pulsars, implying that energetic electron/positron pairs are escaping from the pulsar wind nebulae and radiating in the ambient interstellar medium. It has been suggested that these extended emissions constitute a distinct class of nonthermal sources, termed ``pulsar halos''. In this article, I will review the research progress on pulsar halos and discuss our current understanding on their physics, including the multiwavelength observations, different theoretical models, as well as implications for the origin of cosmic-ray positron excess and Galactic diffuse gamma-ray emission.
\end{abstract}

\ccode{PACS numbers:}

%\tableofcontents

\section{Introduction}
{A pulsar is believed to be a fast-rotating highly magnetized neutron star, from which periodic pulsed electromagnetic radiation is observed. The radiation is powered by the rotational energy of the pulsar as the pulsar gradually slows down. In fact, the radiation only carries away a small fraction of the spin-down power of the pulsar. Most of the spin-down energy is emitted in the form of an ultrarelativistic magnetized wind, composed of electron and positron pairs\cite{Pacini73}. When the pulsar wind encounter the ambient medium, a strong termination shock is generated\cite{Rees74}. Electron/positron pairs in the wind are further accelerated at the shock, forming a bubble of ultrarelativistic nonthermal particles beyond the termination shock. The shocked pairs give rise to the broadband electromagnetic radiation, forming the so-called pulsar wind nebula (PWN)\cite{GS06}. {Some of} PWNe, in particular those powered by young, energetic pulsars, are believed to be efficient particle accelerators. Recent discovery of 1\,PeV($=10^{15}\,$eV) gamma-ray emission from the famous Crab Nebula implies existence of PeV electron accelerator or perhaps 10\,PeV proton accelerator inside the source\cite{LHAASO21_sci}.}

In 2017, the High-Altitude Water Cherenkov Observatory (HAWC)\footnote{  HAWC is a high-energy astrophysical experiment aiming for $0.1-100$\,TeV ($1\,{\rm TeV}=10^{12}\,$eV) gamma-ray and cosmic ray, located on the flanks of the Sierra Negra volcano near Puebla, Mexico at an altitude of 4100 meters. Readers may refer to Ref.~\cite{Miguel14} and the experiment's homepage for more details, \url{https://www.hawc-observatory.org/}.} reported discovery of extended gamma-ray emissions up to about 50\,TeV around two nearby middle-aged pulsars\cite{HAWC17_Geminga}, i.e., the Geminga pulsar and PSR~B0656+14, respectively, well above the background fluctuation. The two-dimensional surface brightness images of the sources revealed by HAWC show a quasi-isotropic, central-bright morphology, extending an angular distance of at least several degrees from the central pulsars. Although the sizes of the sources are not clearly determined, the sources are obviously more extended than the PWNe associated with the two pulsars. Taking Geminga as an example, the physical scale of its PWN is about $0.2-0.3$\,pc as measured in the X-ray band \cite{Caraveo03, Posselt17}, while the TeV emission around it as revealed by HAWC extends at least 20\,pc given a distance of the pulsars to be $200-300\,$pc. The phenomenon can be explained as the inverse Compton (IC) scattering of energetic electron/positron pairs, which were accelerated in pulsar wind nebulae (PWNe) and escape to ambient interstellar medium (ISM), on the cosmic microwave background (CMB) and interstellar infrared (IR) radiation field. In particular, gamma-ray photons above 10\,TeV are supposed to be primarily up-scattered from the CMB since the contribution from the IR background is suppressed by the Klein-Nishina effect. {  Given the typical photon energy of CMB is about $\epsilon=6\times 10^{-4}\,$eV, the relation between the energy of emitting electron/positron $E_e$ and that of the up-scattered photon $E_\gamma$ can be given by $E_\gamma\approx(E_e/m_ec^2)^2\epsilon=25(E_e/100{\rm TeV})^2\,$TeV. Thus,} the continuation of the gamma-ray spectrum up to 50\,TeV implies the presence of electrons/positrons with energies well above 100\,TeV. The extended sources around the two pulsars discovered by HAWC are referred to as ``TeV halos'' and considered as a new class of gamma-ray sources. Since the escaping electron/positron pairs can in principle radiate in a much broader energy bands, we term this kind of sources ``pulsar halos'' in this article.

\begin{figure}
\centering
\includegraphics[width=1\textwidth]{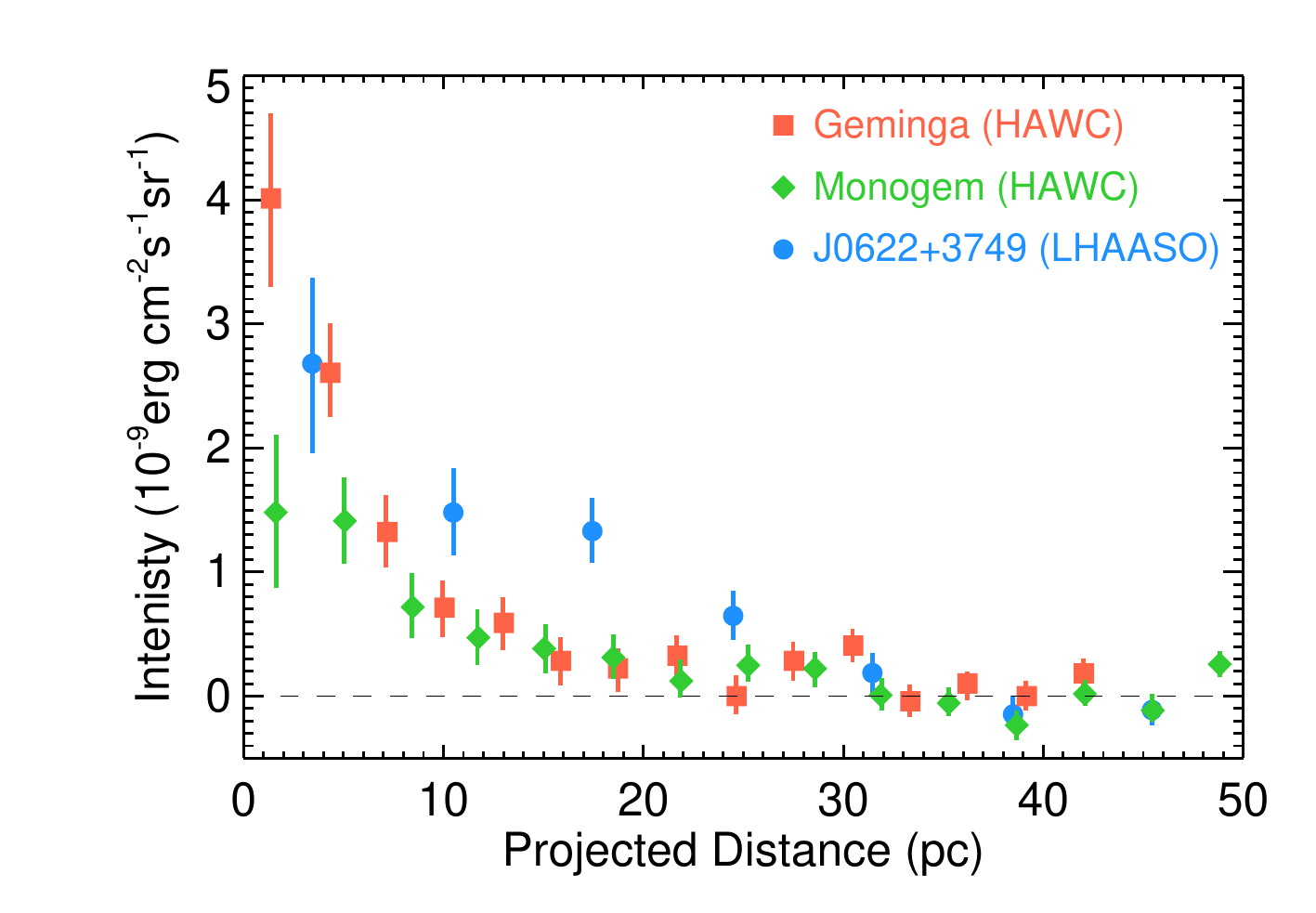}
\caption{Comparison of the TeV gamma-ray intensity profiles of the three identified pulsar halos. Red squares and green diamonds are the measurements by HAWC on Geminga and Monogem respectively between $8-40$\,TeV\cite{HAWC17_Geminga}, while blue circles show the measurement by LHAASO on PSR~J0622+3749 above 25\,TeV\cite{LHAASO21_J0621}.}\label{fig:intensityprofile}
\end{figure}

In general, we may expect the presence of pulsar halos around other pulsars. However, among many candidates of pulsar halos measured by HAWC \cite{HAWC20_3HWC} and the Large High--Altitude Air Shower Observatory (LHAASO)\footnote{  LHAASO is a new generation multi-component instrument, built at 4410 meters of altitude in the Sichuan province of China, aiming to measure gamma-rays from 0.1\,TeV to 1\,PeV and cosmic rays between 1\,TeV and 1\,EeV($=10^{18}\,$eV). More details can be found in Ref.~\cite{LHAASO_WP}.}, only LHAASO~J0621+3755 seems to be another promising pulsar halo\cite{LHAASO21_nat}. The centroid of the source is spatially coincident with a middle-age pulsar PSR~J0622+3749, while there is no other astrophysical counterpart showing up in the region as the possible source of the TeV gamma-ray emission. Also, PSR~J0622+3749 has a similar property to the Geminga pulsar and the Monogem pulsar such as the rotation period, spindown luminosity as shown in Table.~\ref{tab:pulsars}, the characteristic age and the spin-down luminosity. Although the angular extension of the source is not as large as the pulsar halos of Geminga and Monogem, probably due to a much farther distance, the intensity profile is consistent with the diffusion model with a comparable diffusion coefficient to that inferred from HAWC's observation on Geminga and Monogem. A comparison of the TeV gamma-ray intensity profiles of the three pulsar halos is presented in Fig.~\ref{fig:intensityprofile}.

\begin{table*}[bp]
\tbl{Properties of the pulsars with detected pulsar halos. From left to right, each column shows the name of the pulsar, the rotation period ($P$), the time derivative of the period $\dot{P}$, the characteristic age ($\tau_c$), the distance from Earth ($d$), the spin-down luminosity ($L_s$), the ``spin-down flux'' ($L_s/4\pi d^2$). The data is based on ATNF catalogue\cite{Manchester05} unless specified.}
{\begin{tabular}{ccccccc}
\hline\hline
Pulsar & $P$ & $\dot{P}$  & $\tau_c$  & $d$  & $L_s$ & $L_s/4\pi d^2$ \\
~ & (s) & ($10^{-14}$) & (kyr) & (kpc) & ($10^{34} \rm erg/s$) & ($10^{-10}\rm erg/cm^2s$)\\
\hline
PSR~J0633+1746$^a$ & 0.237 & 1.097 & 342 & 0.25$^c$ & 3.2 & 43 \\
PSR~B0656+14$^b$ & 0.385 & 5.494 & 111 & 0.29 & 3.8 & 38\\
PSR~J0622+3749 & 0.333 & 2.542 & 208 & 1.6$^d$ & 2.7 & 0.88\\
\hline
\end{tabular}\label{tab:pulsars}}
{\scriptsize
$^a$Geminga\\
$^b$Monogem\\
$^c$Faherty et al. (2007)\cite{Faherty07}\\
$^d$LHAASO Collaboration (2021)\cite{LHAASO21_J0621}}
\end{table*}

For many other pulsar halo candidates, however, we do not find such a clear association between the extended gamma-ray source and the pulsar. For example, the centroids of the pulsar halo candidate sometimes significantly deviate from the positions of the candidate pulsars \cite{HAWC20_3HWC}, while the angular distance between the centroid of the pulsar halo above 10\,TeV and the related pulsar is supposed to be small due to the rapid cooling of these energetic pairs unless extreme conditions are present \cite{ZhangY21}. Such a deviation might be caused by more complex (but realistic) transport mechanism of escaping leptons, or by additional contribution of other sources in the same region. The former scenario is related to the nature of the interstellar turbulence around the pulsar, which is not fully understood yet and will be further discussed below. The latter scenario is rather common. We can see from the sky maps of the Galactic plane observed by HAWC\cite{HAWC20_3HWC} and by the High Energy Stereoscopic System (HESS)\cite{HESS18_GPS} that many extended sources overlap with each other. There could also exist multiple candidates of the astrophysical counterparts for one extended source. For example, the pulsar halo candidate 3HWC~J1912+103 discovered by HAWC is in spatial coincidence with the source measured by the  HESS, HESS~J1912+101. However, it is suggested to be a supernova remnant (SNR) according to the morphological analysis by Ref.\cite{HESS18_J1912}. A multiwavelength data analysis shows that the source may be understood as the superposition of a pulsar halo and an SNR, with a dominant SNR component of hadronic origin above TeV \cite{ZhangH20}. 

\begin{figure*}
\centering
\includegraphics[width=0.9\textwidth]{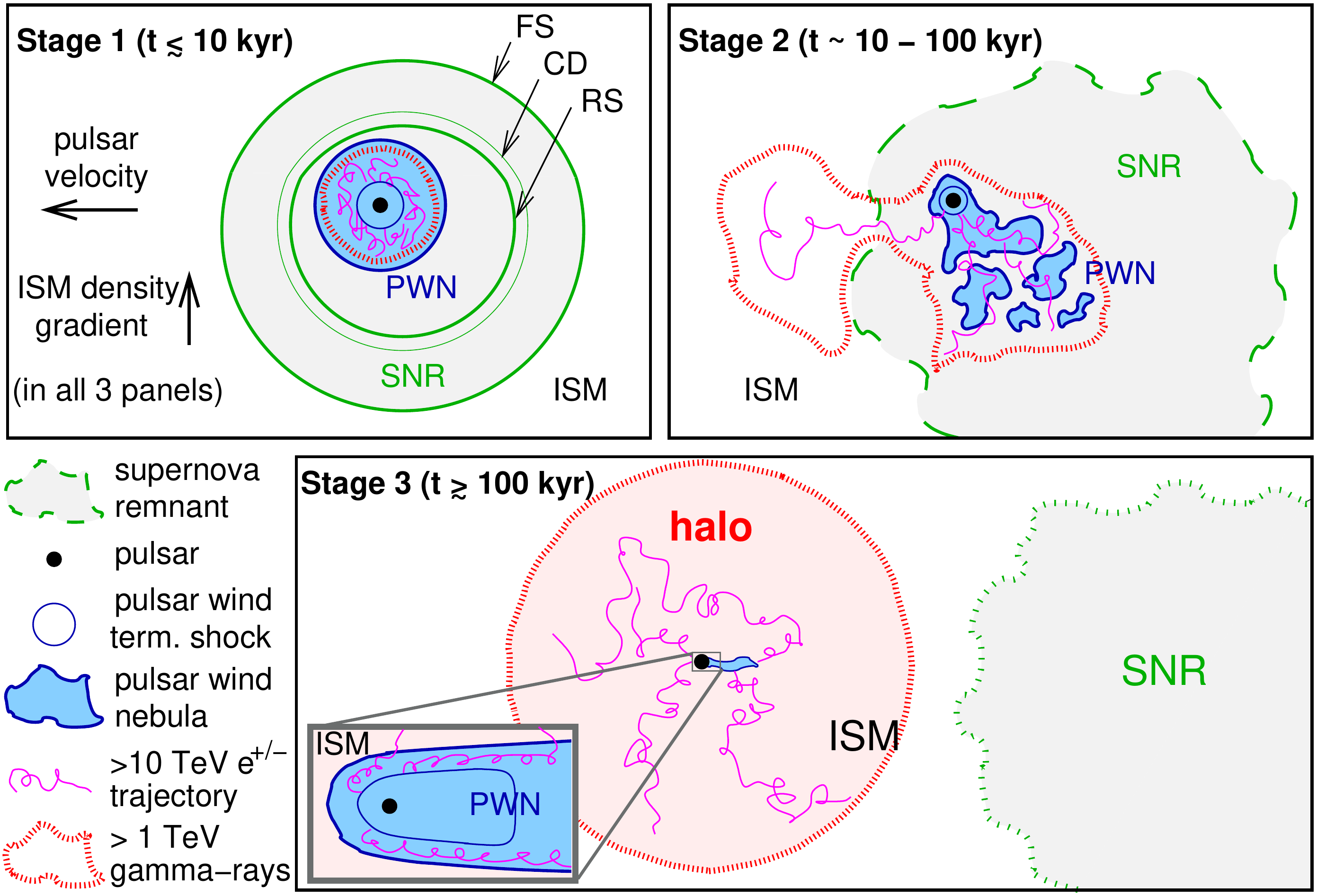}
\caption{A sketch illustration for the 3-Stage evolution of a pulsar system. {  Top Left}: Stage 1 ($t\lesssim 10\,$kyr). The SNR's forward shock (FS) is expanding outwards while the reverse shock (RS) swept through the SN ejecta inwards. The PWNe is expanding in the central of SNR. Accelerated electron/positron pairs are well confined in the PWN; {  Top Right}: Stage 2 ($t\sim 10-100\,$kyr). The RS has crushed (part of) the PWN. Pairs start to escape into the interior of the SNR or perhaps also into the ISM; {  Bottom}: Stage 3 ($t\gtrsim 100\,$kyr). The pulsar has left the faded SNR due to the proper motion. High-energy pairs can efficiently escape into the ISM and form the halo. The figure is taken from Ref.\cite{Giacinti20} with permission from Astronomy \& Astrophysics \copyright ESO. See text for more discussion.}
\label{fig:sketch}
\end{figure*}

Furthermore, a pulsar halo may be confused with a PWN. This is related to the definition of the pulsar halo. Apparently, these two kinds of astrophysical objects are both powered by pulsars and hence physically associated. Ref.\cite{Giacinti20} proposed a 3-stage evolution of a PWN system as shown in Fig.~\ref{fig:sketch}: (1) $t\lesssim 10$\,kyr after the birth of the pulsar. The PWN is expanding inside the SNR and the SNR's reverse shock has not encounter the PWN. The representatives are those plerionic SNR such as Crab Nebula\cite{GS06}, 3C~58\cite{Bocchino01} and G21.5-0.9\cite{Matheson05}. At this stage, the accelerated electrons/positrons are mostly confined inside the PWN; (2) $t\sim 10-100$\,kyr. The reverse shock has (partially) crushed and disrupted the PWN. Some of high-energy pairs can escape the PWN and travel into the SNR or even the ambient ISM. At this stage, the emission of the system may come from both confined and escaped pairs and the morphology of the source could be extended and asymmetric (if the crushing of the reverse shock is asymmetric), such as Vela X\cite{Blondin01a} and HESS~J1825-137\cite{HESS06_PWN}; (3) $t \gtrsim 100\,$kyr (or at least several tens of kyr). The pulsar has travelled beyond the SNR due to the natal kick velocity. Even if not, the SNR has probably faded away and becomes dynamically unimportant. As a result, the pulsar propagates in the ISM and generates a bow-shock PWN if the proper motion is sufficiently fast. At this stage, high-energy pairs can largely escape the PWN and diffuse in the surrounding ISM. Stage 3 corresponds to the cases of Geminga, Monogem and PSR~J0622+3749. Ref.\cite{Giacinti20} defined the system at this stage as the pulsar halo, in which the pulsar no longer dominates the dynamics of the background medium. On the other hand, Ref.\cite{Linden17, Sudoh19} suggested a system to be a pulsar halo if the diffusion is the dominant transport mechanism of pairs. Therefore, a considerable fraction of systems in Stage 2 or even in Stage 1 may be regarded as pulsar halos under this definition. We note that the nature of the sources do not depend on the definition, but the properties of the background medium where the pairs are propagating may be distinct at different stages. For instance, the background medium at Stage 1 and Stage 2 can be more magnetized and more turbulent than that at Stage 3 (i.e., the ISM), so the diffusion coefficients and subsequently the transport mechanisms of particles may differ from each other at different stages. This article will conservatively focus on the sources corresponding to Stage 3. When it comes to Stage 1 and Stage 2, PWNe will be inevitably involved. It would then bring a quite broad topic which is beyond the scope of this article and is also beyond what I can handle for now. So I instead refer readers to previous review papers\cite{GS06, Hester08, Buhler14, Kargaltsev15, Bykov17, Reynolds17, Amato20} on the observations and physics of PWNe. 

Pulsar halo--like sources were actually predicted and investigated in Ref.\cite{Aharonian04} before the discovery of the sources, where the author showed that the morphology or the intensity profile of the extended emission is highly dependent on the diffusion coefficient of the ambient medium. Indeed, by fitting the azimuthal angle--averaged 1D intensity profiles of the pulsar halos of Geminga and Monogem with the isotropic homogeneous diffusion model, Ref.\cite{HAWC17_Geminga} found that the required diffusion coefficient (i.e., $D(E_e)=(4.5\pm 1.2)\times 10^{27}(E_e/100{\rm TeV})^{1/3}\rm cm^2/s$) is 2-3 orders of magnitude smaller than the typical diffusion coefficient of the ISM based on the measurements of local cosmic rays (CRs). However, such a small diffusion coefficient is not the only choice if other particle transport models are considered\cite{Liu19_ani, Recchia21}. Clearly, how escaping pairs propagate around the pulsars is crucial to understand pulsar halos. Therefore, this is still an open question to be addressed in the future. In this article, I will try to provide a comprehensive review on the efforts so far dedicated to understanding the nature of pulsar halos, but I apologize here for I can not exhaust all the relevant papers. The rest of the article is organized as follows: I will describe the multiwavelength observation of pulsar halo in Section 2 which can put constraint on the properties of pulsar halos and particle transport models; as the main part of this article, three main particle transport models for pulsar halos will be summarized and compared in Section 3; in Section 4, I will introduce the potential role of pulsar halos in high-energy astrophysics such as their possible contribution to the cosmic-ray (CR) positron excess and to the Galactic diffuse gamma-ray emission; lastly, a summary of this review and a brief prospect for the future study on this topic will be present in Section 5.

\section{Multiwavelength Observation}
According to the observed broadband emission of PWNe such as Crab Nebula, we can infer that the spectrum of accelerated electron/positron pairs in PWNe spans several orders of magnitude, e.g., from GeV to PeV. Therefore, the spectrum of escaping electrons/positrons is supposed to extend over a wide energy band too, although the spectral shape may be modified due to the energy-dependent escape of particles from PWNe\cite{Bucciantini18}. For instance, 1\,TeV electrons/positrons typically radiate about 40\,GeV gamma-ray photons in a 30\,K IR radiation field via the IC process. Besides, electrons/positrons that are responsible for the multi-TeV emission also radiate in the interstellar magnetic field via the synchrotron radiation, giving rise to radiation at typical energy $E_{\rm syn}\simeq 2(E_e/100{\rm TeV})^2(B/3\mu {\rm G})\,$keV, which is at the X-ray band. Therefore, we may expect a broadband emission from pulsar halos and hence multiwavelength data can bring us additional information on pulsar halos.

\begin{figure*}[htbp]
\centering
\includegraphics[width=0.47\textwidth]{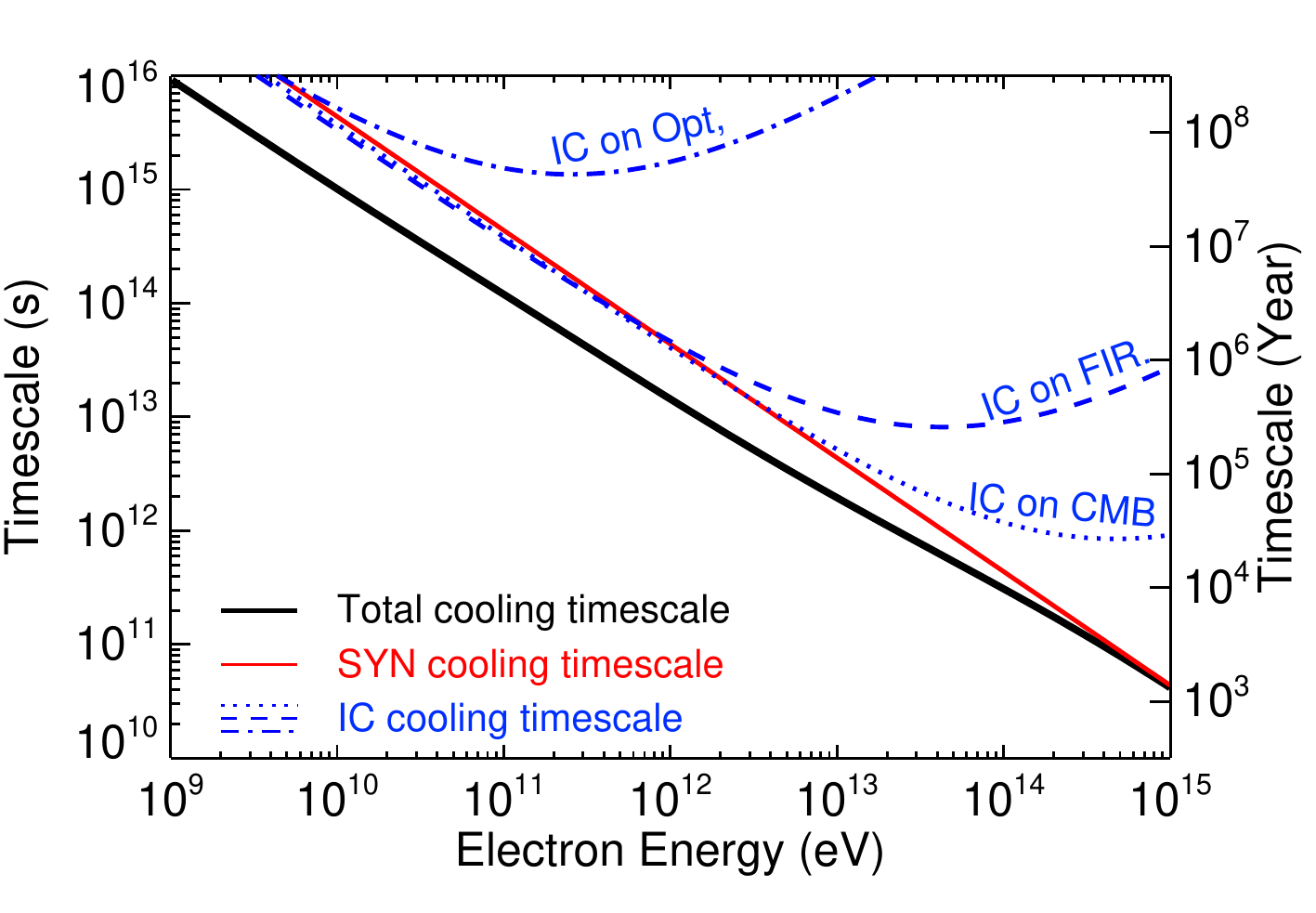}
\includegraphics[width=0.47\textwidth]{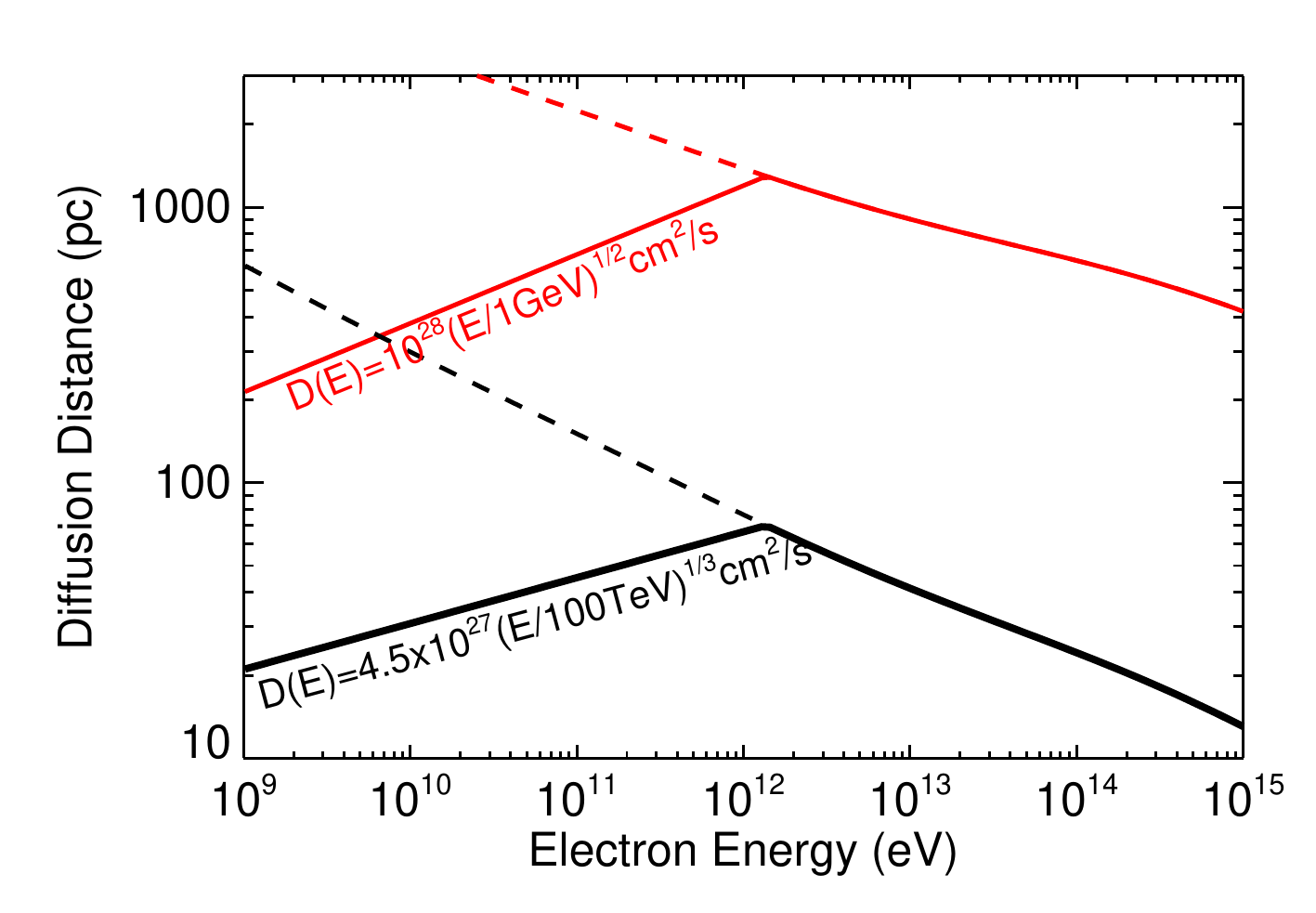}
\caption{{  Left}: Cooling timescale of electrons/positrons in the magnetic field with $B=3\,\mu$G and radiation fields including the CMB, a 30\,K IR radiation field and a 5000\,K optical radiation field with both energy density being $0.3\,\rm eV/cm^3$; {  Right}: Expected diffusion distance of pairs assuming a homogeneous diffusion coefficient with the cooling timescale shown in the left panel and a pulsar age identical to Geminga's characteristic age, i.e., 342\,kyr, for the best-fit diffusion coefficient obtained by HAWC (black) and the standard ISM value (red). Dashed lines show the diffusion distance limited solely by pair cooling.}
\label{fig:cooling}
\end{figure*}

Pulsar halos are spatially extended without clear boundaries. Given the best-fit diffusion coefficient derived from HAWC's observation under the isotropic homogeneous diffusion model, i.e., $D(E_e)=4.5\times 10^{27}(E_e/100{\rm TeV})^{1/3}\rm cm^2s^{-1}$, one may estimate the characteristic size of the halo or the diffusion distance of the pairs by $r_{\rm diff}=2\sqrt{D(E_e){\rm min}[t_{\rm c},t_{\rm age}]}$, where $t_{\rm c}$ is the cooling timescale of electrons/positrons due to synchrotron radiation and the IC radiation and $t_{\rm age}$ is the age of the pulsar (e.g., 342\,kyr for the case of Geminga). From Fig.~\ref{fig:cooling}, we see that injected pairs can distribute over a quite large volume even with the assumption of a suppressed diffusion coefficient over the entire space. For a pulsar distance of $d=200-300\,$pc, the corresponding angular size can reach at least several degree. Therefore instruments with large field of view would be preferable to observe pulsar halos of these nearby pulsars. Otherwise, it'd be difficult to disentangle the background from the extended, diffusive emission by the escaped pairs. 

\subsection{GeV gamma-ray observation}
Pulsar halos are good candidates for Fermi-LAT which operates primarily in an all-sky survey mode and scans the entire sky every three hours. It mainly works in the energy band of $0.3-300$\,GeV. The corresponding energy of pairs that radiate in Fermi-LAT's energy band via the IC scattering in IR radiation field is $\gtrsim 0.1\,$TeV up to a few TeV. Ref.\cite{Xi19} and Ref.\cite{Dimauro19} have separately analysed the Fermi-LAT data around Geminga and Monogem. Both two groups tested many spatial template based on the isotropic diffusion model but the conclusions are different. Ref.\cite{Xi19} reported no significant detection of the pulsar halo by Fermi-LAT while Ref.\cite{Dimauro19} claimed significant detection. 

The major reason for the difference is probably due to different region of interest (ROI) considered in the two works, where Ref.\cite{Xi19} and Ref.\cite{Dimauro19} adopted a $40^\circ \times 40^\circ$ ROI and a $70^\circ \times 70^\circ$ ROI respectively. Another reason is that Ref.\cite{Dimauro19} considered the proper motion of the pulsar when making the template while Ref.\cite{Xi19} did not, while taking into account the proper motion increases the TS value\cite{Dimauro19} by about $30-40$. It may be worth noting that the TS value varies by more than 80 considering different spatial template employed in the analysis, implying that the interstellar emission may affect the conclusion of detection significantly. A more careful study may be needed to clarify in future.

%As further investigated by Ref.\cite{Meng22}, there is an extended weak source to the southwest of Geminga's position in the GeV sky, the centroid of which is approximately on the extension line of the reverse of the Geminga's proper motion. As such, employing a larger ROI can include these counts into the analysis and hence increases the TS value. Similarly, when taking into account of the proper motion, an enhanced emission would appear around the birth place of the pulsar in the obtained spatial template at GeV band \cite{Johannesson19, Dimauro19, ZhangY21}, which is more consistent with the observed image. Nevertheless, it is not clear that whether this weak source is a part of the pulsar halo or not. If not, a stringent upper limit of the GeV flux for Geminga's pulsar halo can be obtained. This implies a hard spectrum of the escaping electrons/positrons spectrum or the spectrum is truncated below certain energy. This essentially shed some lights on the particle acceleration and/or escape processes in Geminga's the bow-shock PWNe. In contrast, if the weak source is demonstrated to be a part of the pulsar halo, it has important implications for the early-time evolution of the PWN.

On the other hand, Ref.\cite{Xi19} and Ref.\cite{Dimauro19} reached a consensus that no significant GeV emission is found around Monogem, but the constraint on the escaping spectrum is not very strong.  Null detection of a GeV halo around PSR~J0622+3749 has also been reported\cite{LHAASO21_J0621}. The halo size is much smaller than that of Geminga so the result is less affected by the background. The obtained GeV flux upper limit has led to a low-energy cutoff or a very hard injection spectrum for the escaping pairs\cite{LHAASO21_J0621}, i.e., harder than $Q(E_e)\equiv dN/dE_edt \propto E_e^{-1}$.

\subsection{X-ray Observation}
As is mentioned at the beginning of this section, electron/positron pairs that are responsible for 10\,TeV gamma-ray emission via the IC scattering off CMB will inevitably radiate X-ray photons via the synchrotron radiation in the interstellar magnetic field. The typical magnetic field strength of ISM is $3-6\,\mu$G, corresponding to an energy density of $0.22-0.9\,\rm eV/cm^3$. This is comparable to the CMB energy density, i.e., $0.26\,\rm eV/cm^3$. Therefore, for pairs with energy $\lesssim 100\,$TeV where the Klein-Nishina effect for CMB is not important, we may expect similar cooling efficiency of pairs via the synchrotron channel and via the IC channel (see Fig.~\ref{fig:cooling}), and hence the pulsar halo's fluxes at 10\,TeV and 1\,keV are supposed to be comparable.

Ref.\cite{Liu19} analysed the archival data of X-ray satellites Chandra and XMM-Newton on Geminga with a total exposure time of 660\,ks and 290\,ks respectively. These two X-ray instruments use the focusing optics and have good sensitivities, but their fields of view are quite limited, which is $16'\times 16'$ for Chandra and $30'\times 30'$ for XMM-Newton. Therefore, when they point to Geminga, they can only observe the very central region of the pulsar halo. 

Without observing the entire halo, the observations of Chandra and XMM-Newton cannot really separate the emission of the pulsar halo from the background. Thus, in the background reduction process, Ref.\cite{Liu19} only removed the emission from background point sources (including that of the Geminga pulsar), Geminga's PWN as well as the instrumental background, but kept the diffuse background including both the Galactic one and the extragalactic one. By doing so, they found that 1D surface brightness profile outside Geminga's bow-shock PWN is flat out to 10' as seen by both instruments. It implies that the observed X-ray emission in the central region of the pulsar halo is dominated by the background. Based on this, a quite low flux limit of $\lesssim 10^{-14}\rm erg/cm^2s$ at $\sim 1\,$keV can be posed for the inner 10' region. They can nevertheless put some constraints on the nature of the pulsar halo.

By fitting the 1D intensity profile measured by HAWC, one can obtain the corresponding synchrotron flux of the inner 10' region from the escaped electron/positron pairs. Under the isotropic diffusion model, Ref.\cite{Liu19} showed that the model flux at 1\,keV is one order of magnitude higher than the upper limit, with an interstellar magnetic field $B=3\,\mu$G and the best-fit diffusion coefficient derived from HAWC's observation. To reconcile the X-ray observation with the TeV gamma-ray observation, one has to reduce the magnetic field down to $B\lesssim 1\mu$\,G and in the meanwhile considering an inhomogeneous (i.e., $r-$dependent) diffusion coefficient, where the detailed upper limit of $B$ depends on the assumed form of the $r-$dependence of the diffusion coefficient. It is unclear yet that why the magnetic field is significantly weaker than the typical value of the ISM. Note that the requirement of the weak magnetic field can be relaxed in the anisotropic diffusion model, which is will further discussed in the next section. 

Strictly speaking, however, the observations by Chandra and XMM-Newton on the inner 10' region of Geminga's pulsar halo do not necessarily represent the property of the entire halo, which extends at least several degree. In principle, these instruments may perform many times pointing to cover the entire halo, although this would be quite expensive. On the other hand, the recently launched X-ray project eROSITA (The extended ROentgen Survey with an Imaging Telescope Array) has a wider field of view than Chandra and XMM-Newton, i.e., $0.9^\circ \times 0.9^\circ$. It is performing a four-year all-sky survey which will finish by the end of year 2023, correspond to a net exposure of 2\,ks of the entire sky except for the ecliptic poles\cite{eROSITA12}. Ref.\cite{LiB22} showed that eROSITA may constrain the magnetic field strength down to $1.4\,\mu$G under the isotropic diffusion model (or Model I, see Section 3.1), if the four-year all-sky survey does not detect an X-ray halo around Geminga. Although this constraint is not as strong as that derived by Ref.\cite{Liu19} with Chandra and XMM-Newton, it can test that whether a weak magnetic field is required for the entire halo region under the model. Of course, the constraint can be improved with more exposure time. The X-ray observation is crucial to understand the nature of pulsar halos as it have much better angular resolution than TeV gamma-ray instrument. With a full coverage of the pulsar halo and a precise measurement on the halo morphology, X-ray instruments may have the power to distinguish different models, as will be discussed in the following section.

\section{Particle Transport Models}
Since pulsar halos are formed by escaped electrons/positrons from PWNe, the key to understand their nature is the spatial distribution of electrons/positrons, which strongly depends on the particle transport mechanism in the ISM around the pulsars. Currently, there are three main models for the particle transport of pulsar halos, and I will introduce and compare them in the following part of this section. 

\subsection{Model I -- Isotropic, Suppressed Diffusion}
\subsubsection{Model I Description}
The most popular model is the isotropic diffusion model (hereafter, we call it Model I for short) with a suppressed diffusion coefficient. {  The evolution of particle density $N_e(E,r,t)$ is governed by
\begin{equation}
\frac{\partial N(E_e,r,t)}{\partial t}=\frac{1}{r^2}\frac{\partial}{\partial r}\left(r^2D(E_e,r)\frac{\partial N}{\partial r} \right)-\frac{\partial}{\partial E_e}\left(\dot{E}_eN\right)+Q(E_e, t)\delta(r),
\end{equation}
where $\delta(r)$ is the Dirac function with $r=0$ being the location of the pulsar. {This term approximates a point-like injection of particles, which is a reasonable assumption given that the halo size is generally much more extended than the PWN.} $D(E_e,r)$ is the diffusion coefficient at a distance $r$ from the pulsar, and $\dot{E}_e$ is the energy loss rate of an electron/positron.}
The simplest version of this model considers a spatially homogeneous slow diffusion zone, i.e., the single--zone diffusion model. This scenario is inconsistent with the measurement on the local secondary-to-primary CR ratio \cite{Aguilar16} and the measurement of the CR electron spectrum \cite{HESS17_electron, Hooper18}. {  An amendment to this scenario} is the two--zone diffusion model raised by Ref.\cite{Hooper17, Fang18, Profumo18, Tang19, Johannesson19}, where the suppressed diffusion zone is considered to be limited only to a radius of $r_0$ (typically a few tens of parsecs) around the pulsar, i.e.,
\begin{equation}
D(E,r)=\left\{
\begin{array}{ll}
D_0(E/100\,{\rm TeV})^{\delta_0}, \quad r<r_0 \\
D_{\rm ISM}(E/100\,{\rm TeV})^\delta_{\rm ISM}, \quad r\geq r_0
\end{array}
\right.
\end{equation}
The solution of particle distribution in the two--zone diffusion model is more complex than that in the single--zone diffusion model, but an analytical solution is still possible as given by Ref.\cite{Tang19}. Alternatively, one can get the solution numerically with various methods, such as did in Ref.\cite{Fang18} with a finite volume method, or as did in Ref.\cite{Profumo18} with a Monte-Carlo algorithm. The results of the three groups generally agree with each other.

Phenomenologically, the small diffusion coefficient in Model I basically arises from the quick decline of the multi-TeV intensity profile at small radius around the pulsar, which corresponds to a pair density distribution steeper than $n\propto r^{-1}$. The latter scaling results from a constant particle injection rate without cooling effect. Since the magnetic field density and radiation field density of the ISM are more or less known, one cannot arbitrarily increase the synchrotron and/or IC cooling rate of the electron/positron pairs to make pairs cool more rapidly. The small diffusion coefficient is therefore invoked to steepen the particle density slope so that pairs can cool before diffuse to the large radius. Note that if we take into account the magnetic field upper limit for Geminga's inner halo obtained by Ref.\cite{Liu19}, the cooling rate of pairs becomes smaller. In this case, a further smaller diffusion coefficient is needed to fit the intensity profile which might violate the so-called Bohm limit. One could  alleviate this tension by employing an increasing function with respect to $r$, instead of a constant value, for the diffusion coefficient within $r_0$ (i.e., $D_0\to D_0(r)$). Such an assumption can lead to a larger gradient in the particle density distribution, and the intensity profile can be reproduced with an appropriate form of $D_0(r)$\cite{Liu19}. This can be regarded as a more complex version of the model.

\subsubsection{Generation of the slow diffusion zone}
 
The key issue in Model I is the physical mechanism of producing the slow diffusion zone. Since particles diffuse isotropically in this model, the magnetic field topology of the ISM around the pulsar is supposedly very chaotic. The small diffusion coefficient implies a significant amplification of the turbulent magnetic component at small scales comparable to the gyroradius of the particle of $r_g=0.04(E/100{\rm TeV})(B/3\mu\rm G)^{-1}\,$pc. A plausible explanation is the CR self-generated turbulence or Alfv{\'e} waves to the gradient of escaped electrons/positrons around the pulsar, i.e., the streaming instability\cite{Skilling71, Skilling75, Bykov13}. 

Such an effect has been studied by several groups. Ref.\cite{Evoli18} considered a 1D diffusion of escaped electron/positron pairs in a flux tube with a transverse radius of $0.1-10$\,pc, and obtained the time and spatial evolution of turbulence excited by those pairs and derive the corresponding diffusion coefficient. They found that the particle diffusion can be suppressed by the self-generated turbulence at early time but will relax to the typical level of the ISM in $\sim 100\,$kyr, i.e., inefficient for middle-aged pulsars unless preferable conditions are considered. Ref.\cite{Mukhopadhyay21} pointed out a numerical error in Ref.\cite{Evoli18} which leads to an overestimation of the non-linear damping rate. They showed that efficient self-confinement of CR with energy up to 10\,TeV can be hold for middle-aged pulsars like Geminga at a distance more than 10\,pc, with a transverse radius of the flux tube $\leq 5\,$pc. Note that, however, the 1D propagation of particles is not consistent with the framework of the isotropic diffusion model, under which the requirement of a slow diffusion zone is derived. The authors further expanded the calculation to a more self-consistent 3D case assuming spherical symmetry of the system. The system volume is significantly increased compared to the 1D case, The CR density is diluted and, as a result, the CR self-confinement can be only hold within 5\,pc from the pulsar. 

Ref.\cite{Fang19} investigated the CR self-confinement effect around Geminga in the most optimistic case in the 3D case with spherical symmetry. They intentionally neglect the cooling of electrons/positrons and the damping of their generated waves to investigate the turbulence excitation under the most preferable assumption. They found that the diffusion coefficient can be at most suppressed by a factor of 5 with respect to the ISM value at 60\,TeV. This is insufficient to explain the suppressed diffusion coefficient. The main reason is that Geminga is too weak to input enough high-energy electrons/positrons to stir the ISM at the late time. \cite{Fang19} also pointed out that considering the proper motion of the pulsar, which is not included in aforementioned two works, may relax the self-confinement of electrons/positrons. This is because the proper motion dilutes the expected particle density around the pulsar especially for those low-energy ones which do not suffer significant cooling. 

Based on these studies, we may conclude that the streaming instability may significantly suppress the diffusion coefficient up to 10\,TeV in the 1D case, but it probably does not work in the 3D case except for the central $\sim 5$\,pc. Note that the self-confinement effect seems to become inefficient for 100\,TeV pairs even if in the 1D case as can be seen in Ref.\cite{Mukhopadhyay21}, due to the small amount of the injected particles and the fast cooling of these extremely energetic pairs. However, the spectra of the three measured pulsar halos all extend to at least several tens of TeV, which are emitted by $\gtrsim 100$\,TeV electrons/positrons. Hence, we probably need to look for other explanations for the suppressed diffusion for 100\,TeV pairs. One possibility might be the non-resonant mode of self-excited turbulence, i.e., the Bell instability\cite{Bell04, Gupta21, Schroer21}. This mechanism requires net current so it would not work if equal amount of electrons and positrons are mixed and injected into the ISM. On the other hand, Ref.\cite{Bucciantini18, Olmi19} suggested that pairs escaping the bow-shock PWN are charge separated, but whether turbulence can be excited before mixing of escaped electrons and positrons due to diffusion is yet to be studied. Ref.\cite{Giacinti18} showed that the electrons and positrons in PWN may not be both accelerated. Either electrons or positrons would be accelerated at the termination shock, depending on the pulsar's polarity. If this is true, the Bell mode may be viable in principle. Alternatively, turbulence may be driven by the external process such as the shock of related SNR of Geminga\cite{Fang19} or by SNR-produced protons\cite{Mukhopadhyay21}. However, there could be a few potential issues in this case. First, it is not sure whether Geminga pulsar (as well as other pulsar with a natal kick velocity) is still located inside the SNR or not, given the proper motion. Besides, unlike those relatively young SNRs with age less than 100\,kyr, whether the SNR-driven turbulence is relaxed or not at the present time also needs a scrutiny. Also, as pointed out by Ref.\cite{Lopez-coto18}, the coherent length of the extrinsic-originated turbulence needs be $\lesssim 5\,$pc in order to be consistent with the apparent quasi-symmetry in the halo's 2D morphology, which may put a constraint on the driving mechanism of the turbulence. 

Lastly, it may be worth noting that the suppressed particle diffusion around pulsars might not be a serious problem for the systems in Stage 1 or Stage 2, because both PWNe and SNRs are more dynamic, and the ambient medium around pulsars could be more magnetized at early time. In addition, the injection luminosity of pairs at early stages can be much higher than that at Stage 3, so the streaming instability might be efficient up to $E_e>100$\,TeV. There have been suggestions\cite{Dimauro20, Breuhaus21, Sudoh21} that some extended gamma-ray sources detected by HESS, HAWC or LHAASO could be explained as pulsar halos of the energetic pulsars in spatial coincidence with them under Model I. I'd suggest to keep in mind that these energetic sources may not be the same as the pulsar halos focused on in this article, since the properties of the ambient medium where particles are propagating could be very different.

\subsection{Model II -- Isotropic, Unsuppressed Diffusion with the Transition from Quasi-ballistic Propagation}
\subsubsection{Model II description}
Isotropic particle diffusion can be approximated as a Brownian random walk, where the mean free path is given by $\lambda(E)=3D(E)/c$. For those freshly injected particles, their transport should not be described as diffusion within a short time period at the beginning, i.e., $t\lesssim \lambda/c$, {  because of insufficient time to get them isotropized.} Otherwise, the average displacement of a particle $\sqrt{4Dt}$ would exceed the distance of particle travelled rectilinearly with the light speed $ct$. This is the problem of the superluminal diffusion, which was addressed in Ref.\cite{Aloisio09} by using the generalized J{\"u}ttner propagator to depict the probability density function of particles injected from a point source after a period of time $t$, {  i.e.,
\begin{equation}
P(E_e,r,t)=\frac{\theta(ct-r)}{(ct)^3Z\left(\frac{c^2t^2}{2\lambda(E_e,t)}\right)\left[1-\left(\frac{r}{ct}\right)^2\right]^2}{\rm exp}\left[-\frac{\frac{c^2t^2}{2\lambda(E_e,t)}}{\sqrt{1-\left(\frac{r}{ct}\right)^2}} \right],
\end{equation}
where $\theta(ct-r)$ is the Heaviside function that truncates the function at $r>ct$ to avoid the superluminal motion, $\lambda(E_e,t)=\int_0^tdt'D(E_e(t'))$ with $E_e(t')$ being the  temporal energy evolution of an electron/positron of energy $E_e$ at $t$, and $Z(y)=4\pi K_1(y)/y$ with $K_1$ being the first-order modified Bessel function. The overall density distribution of electrons/positrons can be obtained by integrating over the contribution from each injection time, i.e.
\begin{equation}
 N_e(E_e,r,t)=\int_0^t P(E_e,t-t',r)Q(E_e,t')dt'.
 \end{equation} }
  It can be demonstrated that for a point source injecting particle with a constant rate $Q$, the particle distribution follows $n(r)=Q/4\pi r^2c$ at $r \ll \lambda$ and $n(r)=Q/4\pi Dr$ at $r\gg \lambda$, i.e, a transition from ballistic propagation to diffusive propagation with an increasing distance from the source. We may see that considering the ballistic propagation would sharpen the gradient of the particle density distribution at the small radius. Such an effect was employed to explain the high-energy spectrum and the gamma-ray intensity profile measured from the Galactic center region\cite{Chernyakova11, Liu16_GC}. Ref.\cite{Prosekin15} further showed that the angular distribution of particles is quite concentrated along the radial direction at small distance from the source, and it would further enhance the central brightness of the source emitted by escaping particles due to the relativistic beaming effect. 

Ref.\cite{Recchia21} pointed out that the steep multi-TeV intensity profiles of pulsar halos can be ascribed to the quasi-ballistic propagation of injected electrons/positrons at small radii (Model II for short). As such, instead of a suppressed diffusion coefficient, a large one is needed to make the mean free path of the emitting electrons/positrons, i.e. $\lambda \simeq 3D(E)/c$, is large enough to cover the inner region of the pulsar halos where the intensity profiles present steep decline. It has been shown that this model can give acceptable fittings to the measured intensity profiles of the three pulsar halos with the standard diffusion coefficient of ISM, although not as good as the fittings in the previous model of the suppressed diffusion. On the other hand, employing a large diffusion coefficient decreases the particle number density around pulsars since particles can travel to a much farther distance from the sources. As a consequence, this model requires a high efficiency of the pulsar's spin-down power being converted to the escaping pair luminosity. While the high conversion efficiency may not be excluded from the first principle, the model is problematic for Geminga because the required efficiency significantly exceeds unity even with a very hard injection spectrum, as also pointed out by Ref.\cite{Bao21}.
  
Nevertheless, Model II may explain the halos around Monogem and PSR~J0622+3749 with possible conversion efficiency. The model does not need to assume an efficient amplification of the interstellar turbulence at small scales around the pulsar, the origin of which is not well determined yet. On the other hand, the isotropic transport of particles in the model still implicitly assumes a chaotic topology of the interstellar magnetic field around the pulsar at the large scale (e.g., the turbulence injection scale). {  This might contradict with the presence of a large diffusion coefficient.} 

\subsubsection{Distinguish Model II from Model I}
Model II would predict a much more extended halo around the pulsar than that in Model I, appearing as a faint tail in the 1D intensity profile. The extended, faint tail may be used to distinguish these two models although it is difficult for detection. Long-term exposure by LHAASO might be available to see this feature. 

On the other hand, the energy-dependent size of the pulsar halo may be another way to distinguish the models. In Model I, the halo size is related to the diffusion distance of pairs. As the higher-energy pairs cool more rapidly, we expect a gradual decrease of the source size (Fig.~\ref{fig:cooling}). Measuring how the halo size changes with energy would shed some lights on the properties of the turbulence. In Model II, the angular distribution of pairs in the ballistic propagation regime are not isotropized and hence the observed emission is only from pairs those are moving towards us. It's been suggested that considering the ballistic motion of injected particles near the source would also shrink the apparent source size above certain energy\cite{Prosekin15, Yang22} in addition to the particle cooling effect. 

It is worth noting that, however, particles most likely move helically along certain magnetic field line, instead of ballistically, at the scale $r<\lambda=3D/c$. This is because the Larmor radius of the particle in typical ISM magnetic field, i.e., $r_L=E/eB=0.04(E/100{\rm TeV})(B/3\mu \rm G)^{-1}\,$pc, must be much smaller than the particle's mean free path $\lambda$. This condition is determined by the prerequisite of Model II, since otherwise the diffusion coefficient should be close to that in the Bohm limit (i.e., resulting in the slow diffusion, similar to Model I). {  To make the quasi-ballistic propagation established, the coherent length of the magnetic field need be shorter than the particle's Larmor radius, which seems not consistent with the typical property of the ISM.} Consequently, we may expect that the angular distribution of particles at small scales has certain basic broadness, which is related to the initial pitch angle of escaping particles with respect to the field line of the ISM in the vicinity of the PWN. Therefore, whether the halo size would be shrunk at high energies or not due to this propagation effect needs a dedicated simulation (cf. Ref.\cite{DeMarco07, Giacinti12_prl}). 

Furthermore, the required injection power of electron/positron pairs in Model II is  orders-of-magnitude larger than that required in Model I (also Model III, see below) given the same spectral shape of escaping pairs. If the high conversion efficiency is common for all the pulsar halos in Model II, it can lead to a different expectation on the diffuse gamma-ray background (DGB) by those unresolved pulsar halos from that in Model I and III would predict, which may provide a clue to the model discrimination. This will be further discussed in Section 4.2.
 
\subsection{Model III -- Anisotropic Diffusion}
\subsubsection{Model III description}
The interstellar turbulence can be sub-Alfv{\'e}nic with Alfv{\'e}nic Mach number $M_A$ less than unity. The interstellar magnetic field then has a mean direction within one coherent length, which is typically $\sim 10-100\,$pc. The isotropic CR transport cannot be hold any more at the scale smaller than the magnetic coherent length scale, because the particle diffusion along the mean magnetic field direction will be much faster than that perpendicular to the mean field direction. According to Ref.\cite{Yan08}, the perpendicular diffusion coefficient $D_\perp$ is suppressed by a factor of $M_A^4$ with respect to the parallel diffusion coefficient $D_\parallel$. {  If we consider the particle distribution in the cylindrical coordinate and denote the mean magnetic field direction as the $z-$axis, the particle density evolution is governed by
\begin{equation}\label{eq:pde}
\frac{\partial N_e}{\partial t}=\frac{1}{r}\frac{\partial}{\partial r}\left(rD_{\perp}\frac{\partial N_e}{\partial r} \right)+D_{\parallel}\frac{\partial^2N_e}{\partial z^2}
-\frac{\partial}{\partial E_e}\left(\dot{E}_eN_e\right)+Q(E_e,t)\delta(r)\delta(z).
\end{equation}}

\begin{figure*}[bp]
\centering
\includegraphics[width=0.45\textwidth]{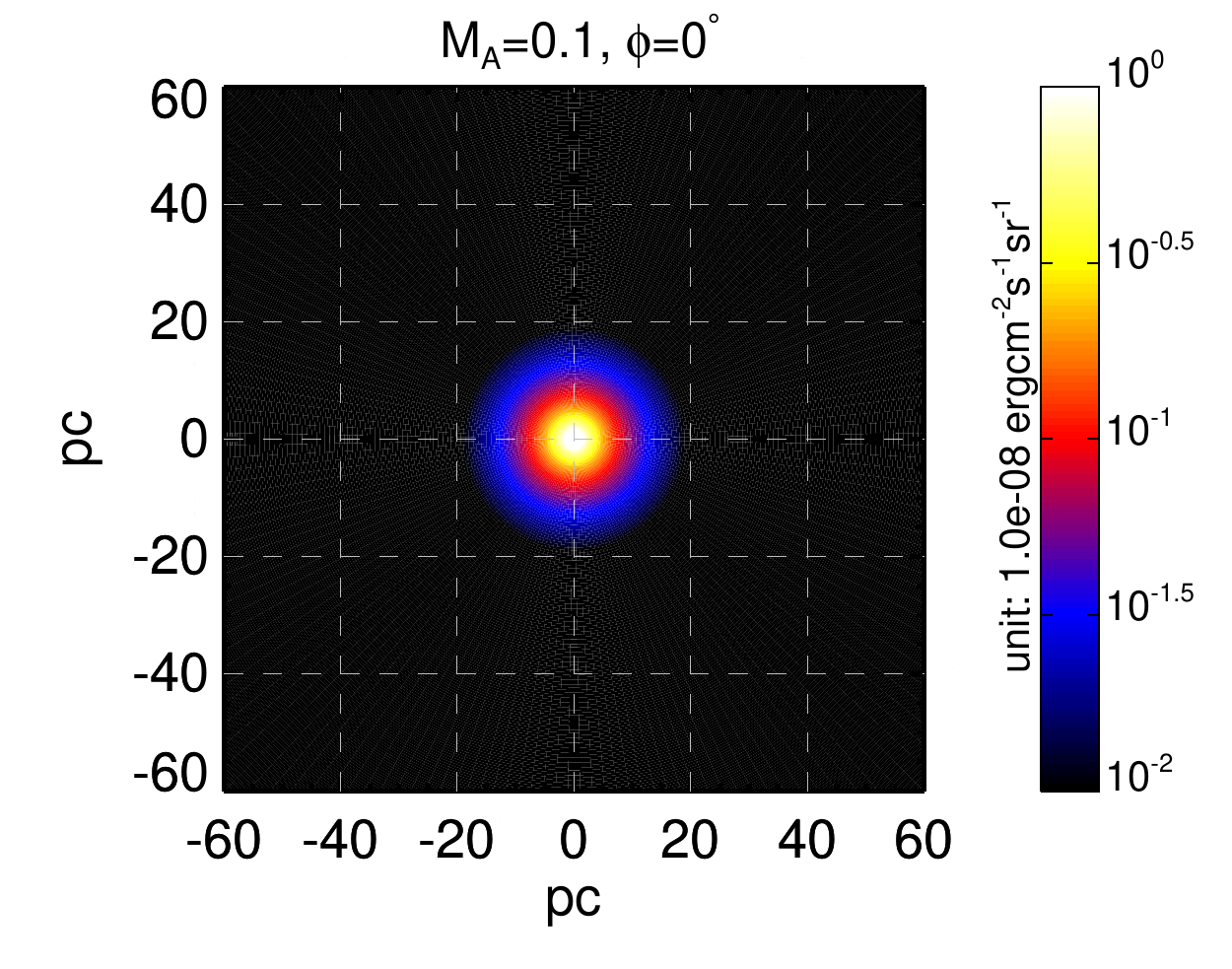}
\includegraphics[width=0.45\textwidth]{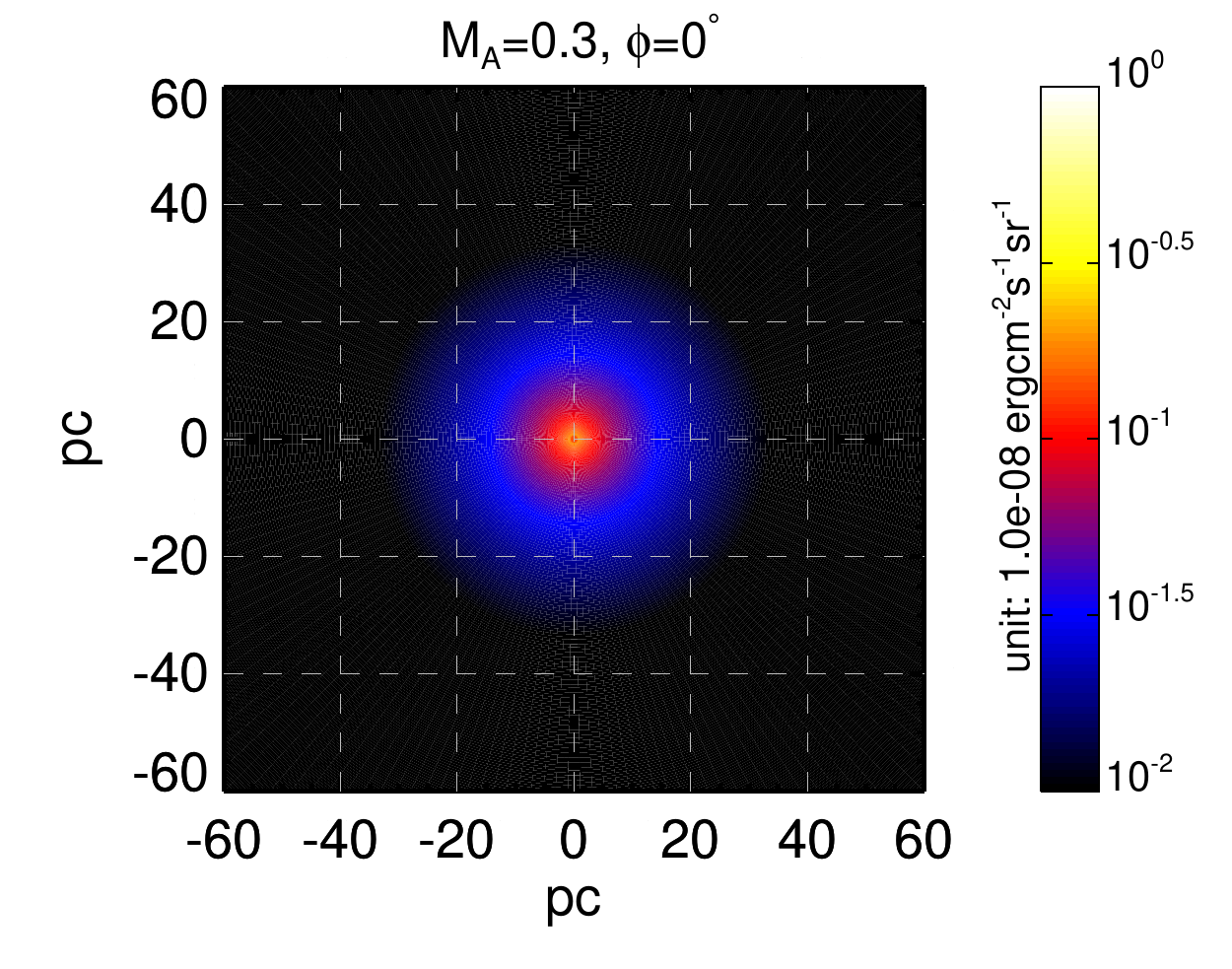}\\
\includegraphics[width=0.45\textwidth]{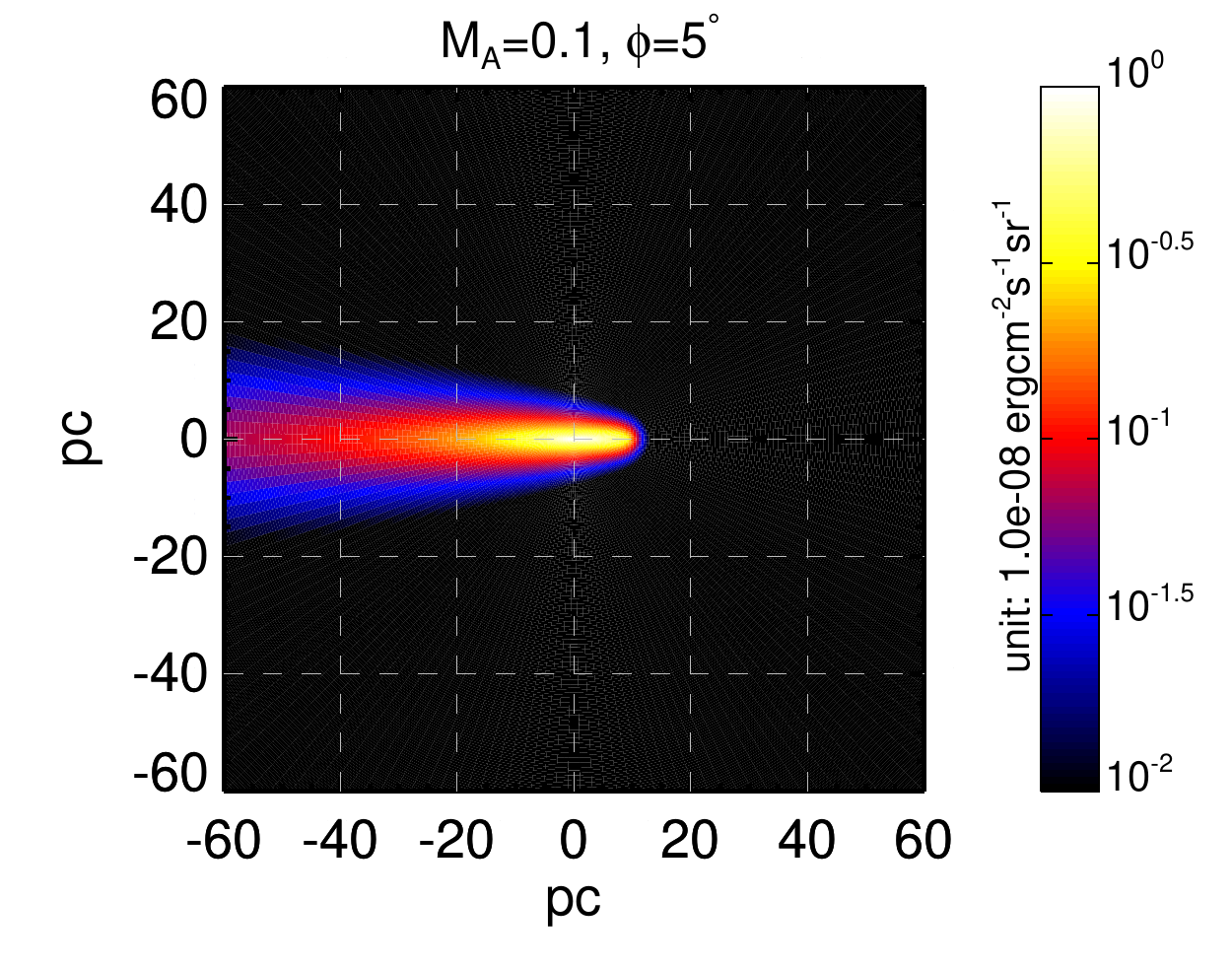}
\includegraphics[width=0.45\textwidth]{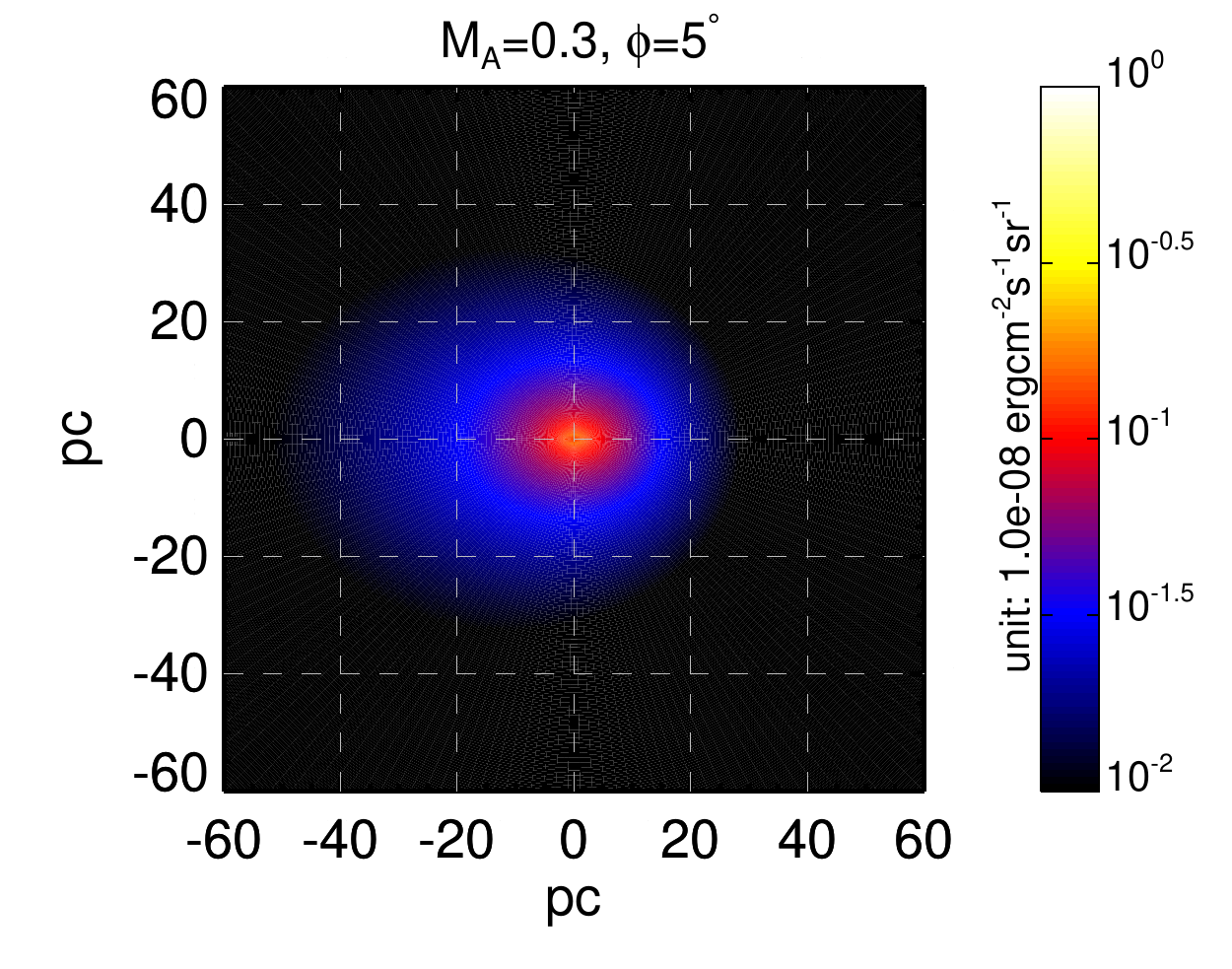}
\caption{Predicted 2D intensity profile in $8-40$\,TeV under Model II with different Alfv{\'e}nic Mach number $M_A$ and different viewing angle $\phi$. This figure is plotted following Fig.2 of Ref.\cite{Liu19_ani}}\label{fig:ani_image}
\end{figure*}

Ref.\cite{Liu19_ani} proposed that the multi-TeV intensity profile of Geminga's pulsar halo can be reproduced under the anisotropic diffusion model (Model III for short) if the mean magnetic field direction in that region is aligned $\lesssim 5^\circ$ with the observer's line of sight (LOS) towards the pulsar. As such, the slow diffusion inferred from HAWC's observation can be naturally explained as the cross-field perpendicular diffusion projected on the plane of the sky, without the need for the amplification of the turbulence. In the meanwhile, the synchrotron flux observed by electrons/positrons are suppressed due to the small pitch angle between the mean magnetic field and the particle moving direction, given that only emission of electrons/positrons moving towards our LOS can be observed. Therefore, the X-ray flux predicted by Model III can be consistent with the X-ray upper limit put by Chandra and XMM-Newton on the central region of the Geminga's halo for a typical ISM magnetic field strength. Such a magnetic field geometry is also consistent with the synchrotron
polarization measurement on the region of $1^\circ$ around Geminga\cite{Gao11}, which shows a very small plane-of-sky magnetic field component. The energy budget requirement of Model III is similar to that in Model I, because both the two models ascribe the steep decline of the intensity profile at the small radius to a slow particle diffusion although the origins of the slow diffusion are different. 
%Note that Model III may not be applicable to the sources around relatively young pulsars (e.g., with age $t\lesssim 100$\,kyr), where the environment may be highly turbulent due to the related SNRs 

\subsubsection{Future Observational Test of Model III}
As shown in Fig.~\ref{fig:ani_image}, the morphology of the halo in Model III highly depends on the inclination angle between the mean magnetic field and the observer's LOS. In principle, the interstellar magnetic field within certain coherent length scale may be oriented in any direction. For the case of Geminga, a small inclination angle is needed to explain the roughly symmetric morphology of the halo and the non-detection of the X-ray emission, but large inclination angles and consequently asymmetric morphology are expected for most other pulsar halos\cite{Liu19_ani, Pedro22, Yan22}. On the other hand, pulsar halos with small inclination angles are more compact and hence more likely being detected, while those with large inclination angles {  are quite extended in the sky and less distinguishable from the Galactic DGB\cite{Yan22}}. Such an observational selection effect might explain the fact that the three detected pulsar halos so far show more or less spherical morphology. Nevertheless, after a long exposure time of HAWC and LHAASO, {it is supposed to be able to see some pulsar halos with asymmetric, elongated morphologies if Model III is right\cite{Yan22}}. Detection of these asymmetric, elongated pulsar halos can serve as a critical test for Model III. 

The X-ray measurement on the synchrotron radiation of the energetic pairs is another way to test Model III. Although all three models predict similar intensity profile at TeV energies (or, we may say that the angular resolution of LHAASO and HAWC are not good enough to be sensitive to the differences), the X-ray instruments generally have much better angular resolutions. Similar to that in the gamma-ray band, the morphology of the X-ray halo predicted in Model III relies on the inclination angle as well. An elongated X-ray halo is a unique feature predicted in Model III. In addition, the ratio of the X-ray flux to gamma-ray flux for a pulsar halo with asymmetric morphology is supposed to be higher than that for the halo with a more or less spherical morphology, provided the same magnetic field strength and radiation density. This can also serve as a test of Model III.

\begin{figure}
\centering
\includegraphics[width=1\textwidth]{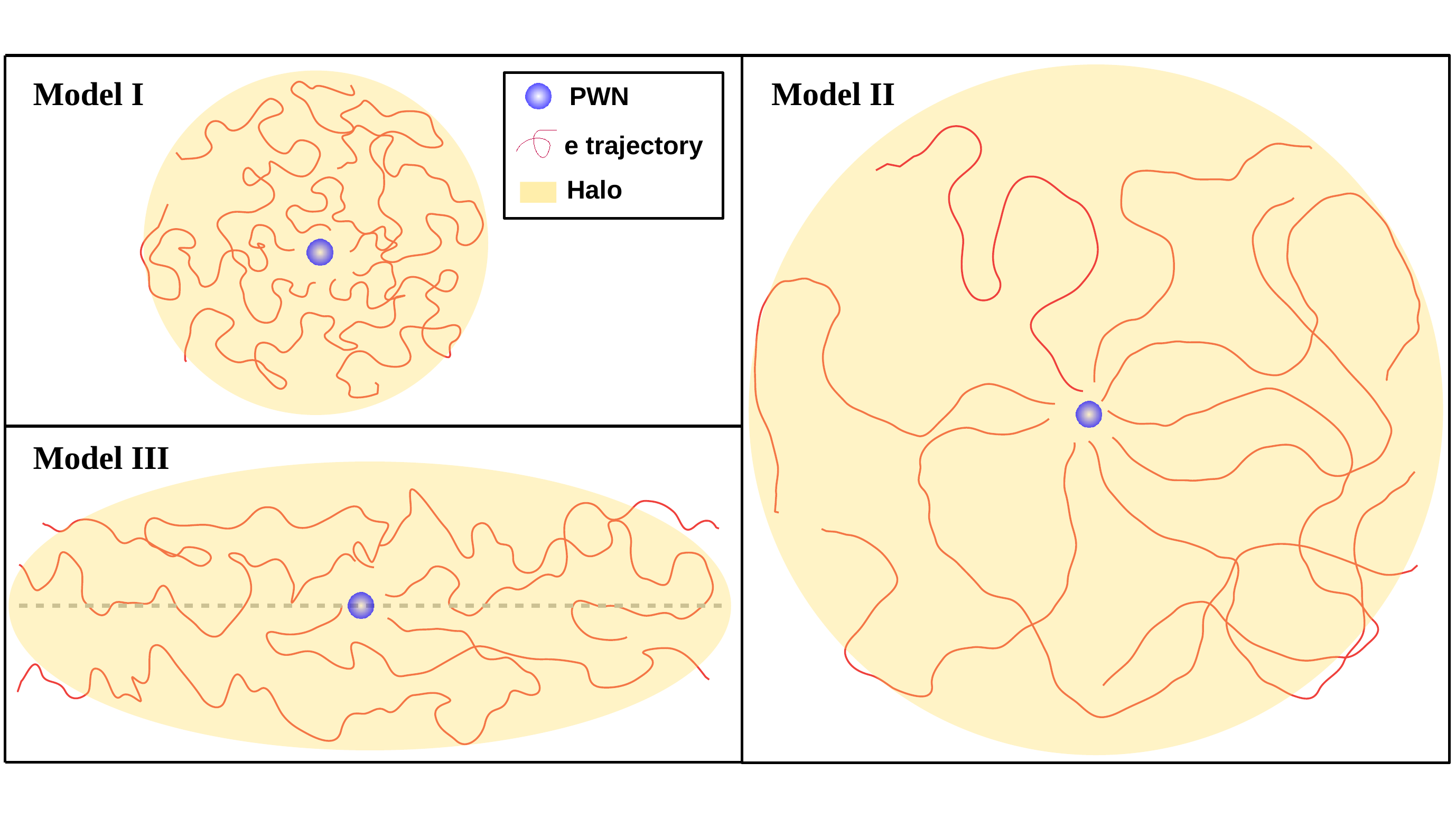}
\caption{An illustrative cartoon of three models. The {blue} disk in the center of each panel means the PWN where pairs are injected. Red curves are trajectories of relativistic pairs. Extended regions colored in {yellow} represent the gamma-ray emission region or the gamma-ray pulsar halo. The grey dashed line in  the panel of Model III represents the mean direction of the interstellar magnetic field around the pulsar. In Model I, pairs are frequently scattered by the turbulence (resulting in a small diffusion coefficient) as they propagate and thus form a compact, quasi-isotropic distribution. In Model II, pairs move quasi-ballisticly at small radius and only get isotropized at large scale due to the large diffusion coefficient. The size of the halo is supposed to be much more extended than that in Model I. In Model III, pairs move preferentially along the mean magnetic field direction given that the turbulent magnetic field is subdominant in the halo. Therefore the halo is extended along the mean field direction but compact in the direction perpendicular to it. However, the observed morphology of the halo highly depends on the observer's viewing angle to the system. See Section 3 for more discussions.}
\label{fig:3models}
\end{figure}

In the end of this section, a comparison of the main properties of the three models is provided in Table.~\ref{tab:models}, {and an illustrative cartoon of the three models is shown in Fig.~\ref{fig:3models}.}

\begin{table*}[htbp]
\tbl{Comparison of three main models for pulsar halos. See Section 3 for detailed discussions.}
{\begin{tabular}{ccccc}
\hline\hline
 Model  &  Diffusion Coefficient &  Magnetic Field & Field Topology & Energy Budget\\
 \hline
I &  $\sim 0.01 D_{\rm ISM}$ & $<1\,\mu$G & Chaotic & $\sim 0.01-0.1L_{\rm s}$ \\
II &  $\sim D_{\rm ISM}$ &  $<1\,\mu$G & Chaotic$^a$ & $\sim L_{\rm s}$ \\
III & $D_\parallel\sim D_{\rm ISM}$, $D_{\perp}\sim 0.01D_{\rm ISM}$ & typical $B_{\rm ISM}$ & Regular$^b$ & $\sim 0.01-0.1L_{\rm s}$ \\
\hline
\end{tabular}\label{tab:models}}
{\scriptsize
$^a$The magnetic field is supposed to be chaotic at the large scale (i.e., the turbulence injection scale) for the isotropic propagation, but the power of the turbulent magnetic field at the small scale that is comparable to particle's Larmor radius should be low for the quasi-ballistic propagation at the inner halo region. This implies the field line is more or less regular at the small scale.\\
$^b$For the halo with apparent spherical morphology, the mean direction of the magnetic field is supposed to be approximately aligned with the observer's LOS.}
\end{table*}

\section{Implications}
\subsection{Contribution to the Positron Excess}
In 2009, PAMELA\cite{PAMELA09} discovered an excess in the CR positron flux from a few tens of GeV up to a few hundred GeV with respect to the prediction based on the hadronic interactions between CR protons and ISM, which is known as the positron excess today. The positron excess is also confirmed with various other instruments such as Fermi-LAT\cite{Fermi12_ee+}, AMS-02\cite{Aguilar13_ee+}, where the latest measurement of AMS-02\cite{Aguilar19_positron} found the energy range of the positron excess extending up to $\lesssim 1\,$TeV at least. The origin of the positron excess could be astrophysical or exotic. 

Nearby pulsars have long been considered as the astrophysical origin of CR positrons\cite{Atoyan95, ZhangL01, Yuksel09}. Detection of the TeV pulsar halo confirms that energetic electrons/positrons are indeed injected into ISM from PWNe. However, if the diffusion coefficient in the intervening space between pulsars and Earth are as small as inferred in Model I, $10-1000\,$GeV positrons that are injected from pulsars would travel too slow to reach Earth during the age of the pulsar, implying that the positron excess may have a more exotic origin such as the annihilation or decay of dark matter particles\cite{HAWC17_Geminga}. Considering the two--zone diffusion model \cite{Hooper17, Fang18, Profumo18, Tang19}, the Geminga pulsar is demonstrated capable of explaining the positron excess solely, given appropriate choices of model parameters. 

It should be pointed out that electrons/positrons which produce emission measured by HAWC and LHAASO are in the energy range of $\sim 10-100\,$TeV, which are not directly related to the positron excess. Positrons with energy relevant to the excess would produce GeV gamma-ray emission via IC scattering off IR and optical radiation field. Therefore, Fermi-LAT's observations of the pulsar halos directly reflect the injection and the propagation of $10-1000$\,GeV positrons. Although Ref.\cite{Xi19} and Ref.\cite{Dimauro19} did not agree on the detection of the GeV gamma-ray halo around Geminga, they reached similar conclusions that Geminga can contribute at most $\sim 10\%$ of the positron excess in Model I, when taking into account the constraint from the GeV observation. Of course, this does not rule out the possibility that other nearby pulsars can make an important contribution to the positron excess. Ref.\cite{Fang19_B1055} proposed that PSR~B1055-52 could be a promising source for the positron excess, if its distance is as small as 90\,pc which is inferred from the dispersion measure. On the other hand, even if a single pulsar could only have a minor contribution to the positron excess, the overall contributions from pulsar population in the Galaxy may still explain the positron excess as shown by Ref.\cite{Manconi20, Evoli21}.

In Model II and Model III, the expected positron flux from a nearby pulsar may be quite different from that of Model I, because the diffusion of positrons can be very rapid. In Model II, the required high conversion efficiency may even lead to an overproduction of the positron flux but note that the flux depends on the spectral shape of the injection positrons too. A harder spectrum or assuming a low-energy cutoff or break may reduce the expected flux. In Model III, the situation is more complex because the intervening ISM between Earth and the pulsar may contain several magnetic field coherences. The exact positron flux relies on the magnetic field topology. Nevertheless, from the current observations we can not confirm nor rule out pulsars as the origin of the positron excess. It'd be more decisive if the CR positron spectrum can be measured to higher energies in the future. The anisotropy of the CR positron would be another critical information to judge the origin. 

\subsection{Contribution to Galactic Diffuse TeV Gamma-ray Emission}
If pulsar halos can be commonly generated around each middle-aged pulsars, we may expect more than one thousand\footnote{I count pulsars with characteristic age between 100\,kyr and 10\,Myr in the ATNF pulsar catalogue\cite{Manchester05}. The reason to set an upper limit of the pulsar age is to exclude millisecond pulsars which have different sources of energy and more complex ambient environment, although I noticed that gamma-ray halos may also form around millisecond pulsars with observable signals \cite{Hooper18_MSP, Hooper21}.} pulsar halos in our Galaxy with an overall spin-down luminosity of about $4\times10^{37}\,$erg/s, noting that this number could be several times greater since there could be off-beamed pulsars which are invisible to us\cite{Linden17}. Most of these halos are currently unresolvable given their small spin-down luminosities and hence the collective emission of these escaped pairs would contribute to DGB, similar to the prediction in Ref.~\cite{Aharonian00}. By simulating a population of sub-threshold pulsar halos, Ref.\cite{Linden18} found that their contribution to TeV DGB may be more important than the $\pi^0$ emission from interactions between CR hadrons and ISM. The predicted intensity and spectrum of the diffuse emission from pulsar halos can match the measurement of Milagro\cite{Milagro05} and ARGO-YBJ\cite{ARGO15_diffuse}. Also, Tibet AS+MD array recently reported discovery of sub-PeV DGB\cite{Asgamma21_diffuse} and it has been shown that contribution from unresolved sources may be required to explain a fraction of this emission\cite{Liu21_diffuse, Fang21_diffuse}. Besides, Ref.\cite{Vecchiotti21} proposed that unresolved pulsar halos (including unresolved PWNe of younger pulsars) may cause the gradual hardening of the diffuse gamma-ray spectrum towards the Galactic center as observed by Fermi-LAT at sub-TeV energies. 

Note that different models for pulsar halos may lead to different contribution level to the TeV DGB. In Model I, we would expect most pairs with energy higher than TeV to deposit their energies in the Galactic disk since these energetic pairs can travel only several tens of parsecs before before cooling, as shown in Fig.~\ref{fig:cooling} with the black curve. Even considering the two-zone diffusion, we may still expect most of their energies lost in the inner slow diffusion zone. On the contrary, we see from the same figure that pairs can diffuse to $r_{\rm diff}\approx 0.6(E_e/100{\rm TeV})^{1/4}\,$kpc before cooling with the standard ISM diffusion coefficient. Usually, we restrain the analysis of the Galactic DGB within a Galactic latitude of $|b|<5^\circ$, implying that we could miss a considerable fraction of the emission of pairs injected from sources within a distance of $d\simeq r_{\rm diff}/\tan 5^\circ\approx 7(E_e/100{\rm TeV})^{1/4}\,$kpc. On the other hand, pairs injected by those comparatively nearby sources can form a DGB at intermediate latitude, i.e., $|b|\approx 35^\circ (r_{\rm diff}/0.6\,{\rm kpc})(d/1\,\rm kpc)^{-1}$. In Model III, although the injected pairs diffuse primarily along the local magnetic field line in the ISM around pulsars, they will diffuse isotropically at the scale larger than the coherent length. Also, the local magnetic field direction in the ISM around different pulsars can be very different. As a result, the global transport of the injected pairs in the Galaxy and hence their spatial distribution would be similar to that in Model II. Here, I'd like to caveat that for Model II and III the influence of the regular component of the Galactic magnetic field might be non-negligible when considering the transport of pairs at the kpc scale (i.e., for pairs of $\gtrsim$TeV energies), which is yet to be investigated. 

The power conversion efficiency of pulsars also plays an important role here. Both Model I and Model III generally require a factor of $\sim 0.01-0.1$ (depending on the assumed slope of the injection pair spectrum) for the efficiency  to explain the measured pulsar halos\footnote{The conversion efficiency also depends on the spectral shape of the injected pairs. But given the same spectral shape, Model I and III require a much smaller conversion efficiency (or pair energy budget) than Model II in order to explain the three detected pulsar halos.}, whereas Model II requires the efficiency close to unity. If such a difference can be generalized to all the pulsar halos, we may conclude that Model III predicts the lowest intensity level of the DGB at the Galactic plane among the three models, because pairs will distribute over a larger volume than that in Model I, while the injection amount of pairs is less than that of Model II. It is not straightforward to judge which of Model I and Model II can give rise to a higher Galactic DGB intensity. The diffusion length of pairs in Model II is about one-order-of-magnitude larger than that in Model I, making its pair distribution less confined in the Galactic disk, but the much higher conversion efficiency in Model II may compensate the larger volume that pairs distribute. Nevertheless, the difference of the DGB intensity at intermediate latitudes may be pronounced, although the DGB at intermediate latitudes may be weak in all three models and undistinguishable by current instruments. In addition, some other factors such as the dark matter decay\cite{Neronov20_diffuse_lso} and CR electrons\cite{HESS17_electron} might also contaminate the DGB at intermediate latitudes. A detailed investigation is needed to quantify these issues. 

Currently, the DGB is not considered in the data analysis of TeV-PeV gamma-ray instruments. We may envisage that although this postulated DGB from pulsar halos, if truly exists, probably do not affect the observation of bright sources such as the first 12 ultrahigh-energy gamma-ray sources measured by LHAASO \cite{LHAASO21_nat}, it may potentially cause problems to the analyses of those extended and comparatively weak sources. For example, based on the calculation of Ref.\cite{Linden18}, the 20\,TeV intensity of the pulsar-halo-induced DGB is about $2\times 10^{-7}\rm \,GeV~cm^{-2}s^{-1}sr^{-1}$, and we may expect its contribution inside a circular region of radius $\theta$ being $F_{\rm DGB}(20\,\rm TeV)\simeq 3\times 10^{-13}(\theta/1^\circ)^2\rm erg\, cm^{-2}s^{-1}$. The sensitivity of LHAASO after 1-year exposure on an extended source with size $\theta$ can be inferred from the point-source sensitivity by $S_{\rm ext}=\left[1+(\theta/\theta_{\rm PSF})^2\right]^{1/2}S_{\rm ps}\approx 7.5\times 10^{-13}(\theta/1^\circ)\,\rm erg\,cm^{-2}s^{-1}$ at 20\,TeV for $\theta>\theta_{\rm PSF}$, where $S_{\rm ps}\simeq 3\times 10^{-13}\rm erg\,cm^{-2}s^{-1}$ and $\theta_{\rm PSF}\approx 0.4^\circ$ are the point-source sensitivity and the size of point-spread-function of LHAASO-KM2A\cite{LHAASO_WP} at 20\,TeV respectively. Note that the sensitivity of LHAASO given here is obtained on the CR background, without considering the DGB. We can see that the contamination of DGB could be already important for those sources with marginal detection after 1-year exposure of LHAASO-KM2A if the size is $\gtrsim 1^\circ$. Of course, LHAASO may also resolve many pulsar halos from the DGB and lower the level of the expected DGB. However, in the case of Model II or Model III, the pulsar halos could be quite extended and overlap with each other because of the fast diffusion of injected pairs, making it very hard to remove them from DGB. In other words, the pulsar-halo-induced DGB may raise the detection threshold of LHAASO or other instruments for those weak, extended sources. If so, a careful evaluation on the pulsar-halo-induced background is needed in order to detect and extract correct information of those weak, extended sources. 

\section{Conclusion}
Pulsar halos are a new class of very-high-energy gamma-ray sources discovered recently around middle-aged pulsars. It is widely believed that their TeV emissions are produced by the inverse Compton radiation of electron/positron pairs escaped from PWNe. However, the particle transport mechanism in the surrounding ISM is not fully determined yet. There are mainly three models able to explain the observed feature, but they have different predictions which can be utilized to distinguish them. To facilitate our understanding on the nature of the sources, we need multiwavelength coverage of the sources as well as searching for a larger sample of the sources. Pulsar halos may also contribute a non-negligible gamma-ray background in the Galactic plane which may play an important role in very-high-energy and ultra-high-energy gamma-ray astronomy. From theoretical point of views, pulsar halos are related to many intriguing issues of modern astrophysics, such as the origin of CRs, the transport mechanism of CRs in ISM, the properties of the interstellar turbulence, the particle acceleration and escape mechanisms in PWNe, the pair multiplicity of pulsars. The future observations with of eROSITA, Cherenkov Telescope Array, LHAASO, SWGO (Southern Wide-field Gamma-ray Observatory) may provide critical information. {  At the finalizing stage of this article, I noticed another review of pulsar halos published online\cite{Lopez-coto22}, with the some contents complementary to this article. Readers may refer to it for a more comprehensive understanding on this topic.}

\section*{Acknowledgments}
The author thank the anonymous referee for the constructive suggestions. The author is grateful to Xiang-Yu Wang for the careful reading and useful suggestions, Gwenael Giacinti for the inspiring discussion on the ballistic v.s. helical motion of particles, Hao Zhou for the valulable discussion on HAWC's data, Songzhan Chen for the help to understand LHAASO's performance, Shao-Qiang Xi for sharing his insight into the Fermi-LAT's observation on pulsar halos, and Kun Fang and Yiwei Bao for interesting discussions on the CR self-generated instabilities. This work is supported by NSFC grant No.~U2031105.

\bibliographystyle{naturemag}
\bibliography{ws-ijmpa}

%\appendix

%\section{Appendices}

%\begin{thebibliography}{000} %for 3 digits
%\begin{thebibliography}{00}  %for 2 digits
%\begin{thebibliography}{0}    %for 1 digit

%%journal paper
%\bibitem{jpap} R. Loren and D. B. Benson, {\it J. Comput. System Sci.} {  27}, 400 (1983).

%%collaboration
%\bibitem{colla} OPAL Collab. (G. Abbiendi {\it et al}.),
%{\it Eur. J. Phys. C\/} {  11}, 217 (1999).

%%normal book (authors)
%\bibitem{autbk} R. Loren and D. B. Benson, {\it Introduction to String
%Field Theory}, 2nd edn. (Springer-Verlag, New York, 1999).

%\end{thebibliography}
\end{document}